# Dynamics of Meaning

Gary Gindler

e-mail address: ggindler@yahoo.com

(Dated: December 21, 2020)

A formal theory of meaning (the process of knowledge accumulation) as multiplicative chaos is proposed. The epistemological process is understood as the process of subjective extraction of some knowledge from the incoming information. The concepts of nonsense are introduced as a meaning that has a minimum value equal to one and the level of intelligence as a geometric mean of the cumulative meaning. The thesis of the multiplicativity of meaning, its polymorphism is substantiated, and numerous examples from world history are provided. By analogy with classical thermodynamics, three laws of thermodynamics of meaning are postulated. Estimates of the cumulative meaning when the comprehension of information (multiplicative cascade) is a random process with given statistical characteristics are carried out.

## 1. Introduction

Almost 100 years have passed since the release of the book with the striking title "The Meaning of Meaning" [1], in which authors Ogden and Richards first proposed some formalism when describing such a complicated concept as "meaning." Their approach's success is based on the fact that from all the variety of problems associated with human consciousness, they singled out only one problem – the question of meaning, leaving aside a significant number of other challenges.

The famous "Ogden-Richards triangle," as it turned out later, could be traced back to the works of Aristotle. Of course, this does not diminish Ogden and Richards' achievements but only emphasizes that the problem of consciousness in general, and meaning in particular, has been known for at least two and a half millennia. This topic's interest is so great that practically all well-known philosophers and scientists have not ignored it. Here, we note the works of only those who are impossible not to mention – Descartes [2] and Penrose [3].

Each branch of science builds its models of consciousness and its models for extracting meaning from the information received. Biologists concentrate on the chemical reactions in the human brain; mathematicians make informational models of the brain; psychologists and philosophers use their specific methods that are poorly understood by representatives of the "exact sciences." The theory of "quantum consciousness" [4] and, of course, its criticism [5] has become popular now. A common feature of these studies is that they are directed to the micro-world; that is, they try to explain consciousness in terms of the dynamics of the "elementary particles" of the brain – neurons.

At the same time, the question remains about the work of the brain, not at the microlevel, but at the macro-level – at the level of an ensemble of a colossal number of certain "building blocks," which, presumably, reside in the human brain, and whose microdynamics provide processing, comprehension, and storage of incoming information. If "building blocks" are understood as neurons, then there are about a billion of them in the human brain – a large enough number to speak of a statistical ensemble. Similar problems are posed in statistical physics and thermodynamics, which describe the macrostates of a vast number of individual objects, of which surprisingly little is known of the internal structures.

This phenomenological work is an example of the Copenhagen school of quantum mechanics, the slogan of which is "*Shut up and compute*," without going into fascinating, but so far inaccessible to our understanding, the actual mechanisms of the brain at the micro-level. In essence, the Copenhagen school's slogan is Descartes' modified slogan: "*Cogito, ergo sum, ergo computo*."

In this paper, many of the exciting things found in discussions of consciousness from a psychological or neuroscience point of view will be deliberately avoided. For example, we will not pay attention to illusions, mental disturbances, mental disorders, self-consciousness, or the ever-popular world of unconsciousness research. We single out just one segment of consciousness – its ability to assign meanings to external information.

The article has the following structure. All mathematical calculations are given in the Appendices, and the central part of the article is deliberately devoid of any formulas so that the paper could be accessible to everyone, not just physicists.

## 2. Information and meaning

Words (meaning both written and spoken language) are a reflection of a person's thoughts. Thus, they are carriers of meaning – but they do not mean as such. It is known that words are also a carrier of information. Still, since information and meaning are different categories, it should be recognized that words have a double load – both informational (meaning syntactic information, Shannon's information) and semantic (meaning semantic information). The recent review [6] can serve as an excellent introduction to the terminology of the issue.

For example, the phrase that begins the U.S. Constitution – "We the people of the United States" contains 34 bytes of information, but in Russian, the same phrase – «Мы, народ Соединенных Штатов» – carries a smaller 28 bytes.

Nevertheless, it is evident that, despite the difference in the amount of information, the two phrases above have the same meaning. Thus, semantic information – meaning – is polymorphic, that is, the same meaning can be presented using different syntactic information streams. In contrast, Shannon's information is polysemantic; that is, different people can derive different meanings from the same information flow. Moreover, often a person extracts different meanings from the same information at different points in time.

In practice, the process of understanding, that is, extracting meaning, is actually two parallel processes. A person analyzes the whole using its parts (for example, reads this article using familiar words), and at the same time, analyzes the parts using the whole (assigns a meaning to individual words that correspond to the context of the entire article). It is impossible to "break" this epistemological circle; therefore, the extraction of meaning is always a subjective process. In other words, the meaning is not the sum of the components; that is, it is not merely the sum of the meanings of individual words in a sentence.

*Meaning is a function of subjective perception of objective information.*

In other words, the non-trivial meaning is that part of the interpretation of objective information that is subjectively assessed by a person as useful, necessary, having some value and significance.

Different people perceive different meanings from the same information (understood traditionally, in a purely mathematical sense), including that trivial form called nonsense. That is, for someone, a particular meaning has a high value, and for someone else – a low one. It shows the objectivity of information and the subjectivity of meaning.

*From a philosophical point of view, information is an objective measure of form, and meaning is a subjective measure of substance.*

### 3. Mathematical properties of meaning

It should be noted that even if the text is a complete gibberish, nonsense, a random set of symbols, then this is its meaning, so that (on an individual level) the meaning never vanishes.

Another argument for the fact that meaning never vanishes (at least at the individual level) is meta-information. Indeed, any information is accompanied by its companion – meta-information – information about itself (what is the source of this information, when it was transmitted, when it was received, what is its size, etc.)

For example, information that some information exists is binary meta-information. Even when the information itself does not make sense from one individual's perspective (has a meaning of nonsense), the meta-information about information – no matter how meaningless it is – exists. That gives reason to assert that some non-zero meaning (which is carried in this case by meta-information) exists as well. Let us assign the nonsense a meaning equals to 1.

A direct consequence of this is that when the same information is received again, a person still extracts some non-trivial meaning.

If the concept of information is additive (the size of the sum of two files with sizes of 1Kb and 2Kb is exactly 3Kb), then in this paper, we postulate that the concept of meaning is multiplicative. For example, if a person has extracted some meanings from these two files, then the whole meaning will be a function of the product of these two meanings.

The multiplicative nature of meaning leads to the fact that the "level of intelligence" of a person must be determined not through the arithmetic mean of the meanings (knowledge) received by him, but as a geometric mean (see Appendix I).

The fact is that if the arithmetic average is well suited for objects that are to some extent independent of each other, the geometric mean is more suitable for describing dependent quantities. Indeed, the meaning of what is read today in most cases depends on what was read and understood by a person yesterday, that is, on the level of education of the individual. The geometric mean correctly describes the well-known accumulation effect – the more educated a person is, the easier he or she assimilates new information (more precisely, the easier it is for the person to extract new meaning from new information).

To a certain extent, the process of accumulating meanings resembles the process of accumulating wealth – the return on investment over several years is the product of the returns for each year individually.

Besides, the geometric mean has an interesting effect – it downplays extreme values. In fact, everyone knows cases when an individual became a brilliant specialist in his field (in our terminology, he extracted an exceptionally great meaning, knowledge, and skills, in some business). Nevertheless, suppose his level of education in all other areas is low. In that case, the subjective assessment of such a person by those around him will also be low – in full accordance with the formula of the geometric mean.

In the opposite case – when a person has an adequate level of knowledge in many areas, but in one case is a complete layman (for example, does not know how to cook). As a rule, this shortcoming is "forgiven" for such a person, in full conformity with the geometric mean property of not attaching importance to the rules' exceptions.

As a non-linear mathematical operation, meaning extraction is a non-linear process. This process resembles a holographic one, in which two waves interact non-linearly with each other on a photographic plate ("multiply" by each other) and thereby create on it not only amplitude but also a phase (spatial) impression. The image on the photographic plate is not concentrated at any specific points (pixels). A three-dimensional image is "sprayed" onto a two-dimensional plate in such a way that each part of it becomes an information carrier. The mechanism of adding new meaning to the existing one works approximately according to the same holographic principle. The idea of "holographic consciousness" was first proposed by Pribram [7], who, however, did not pay attention to the essential fact that holography is fundamentally a non-linear process.

The multiplicativity of meaning also leads to the fact that the knowledge gained, the realized meaning, becomes a part of the meaning accumulated by a person, and it is impossible to somehow "separate" a fragment of the meaning already received. Therefore, as soon as new knowledge penetrates consciousness, it "dissolves" in the already accumulated meaning (formally, the "old meaning" is multiplied by a certain amount, greater than 1) and becomes its integral, inseparable part. The just received fragment of meaning ceases to be a fragment, ceases to be a part of the whole – it becomes whole, "diffuses" throughout the entire consciousness.

The non-commutativity of individual fragments of the meaning is also intuitively clear. For example, let us compare two people who decide to fly an airplane. One person decided to fly in the sky before he completed flight school, and another only decided to fly after first completing flight school. The extraction of meaning by one of these two pilots will bring many surprises.

Another example of the non-commutative nature of meaning is visual arts. Consider two people contemplating Raphael's Sistine Madonna. One person learned that the depicted Pope Sixtus II had six fingers on his right hand only after visiting the Dresden Gallery, and another knew about this even before visiting. There is no doubt that these people will derive completely different meanings from what they saw.

There are two categories of people – one type evaluates the fine arts only by its technical characteristics, and the other – only based on those emotions and those evoked associations. Both are quite acceptable, but due to the property of non-commutativity, both categories of people will only benefit if they listen to an alternative point of view. Obviously, both categories will benefit from the knowledge that Raphael belonged to the Gnostics, who, of all numbers, the number six was particularly honored.

*The result of the extraction of meanings depends on single episodes of cognition and the sequence of their receipt.*

How diverse are the meanings? How big is the universe of meanings? First, different people can derive different meanings from the same information. Secondly, even a simple rearrangement of the sequence of obtaining two meanings leads to a different result. Thirdly, the human apparatus for operations on meanings – the brain – is imperfect; it is human nature to forget.

Therefore, the totality of all kinds of perturbations of meanings in human society is enormous. The probability of finding two people with the same baggage of meanings may be zero. The meanings are probably even more unique than fingerprints. As shown in Appendix II, cognition trajectories "diverge" exponentially overtime for any two people.

A prime example is the BBC broadcast on the eve of the Allied invasion of Normandy in 1944. On June 1, 1944, the BBC aired the first three lines of Paul Verlaine's poem "Autumn Song":

*Les sanglots longs*
*Des violons*
*De l'automne.*

Then, on June 5, a few hours before the invasion, what went down in history as "D-Day," the following three lines were transmitted:

*Blessent mon coeur*
*D'une langueur*

*Monotone.*

Here is the translation of Verlaine's poem into English by Arthur Simons:

*When a sighing begins
In the violins
Of the autumn – song.*

*My heart is drowned
In the slow sound
Languorous and long.*

Different people have drawn entirely different meanings from these short messages. For the uninitiated, it was just poetry. For the French Resistance fighters, it was a signal that an invasion was imminent, and it was time to complete their assignment. For the Third Reich intelligence officers, who knew about this signal, it meant almost the same as for the Resistance fighters. Intelligence reported everything to Berlin, but Berlin appropriated this information with the meaning of nonsense and ignored it.

This episode illustrates that the extraction of new meaning is based on already accumulated meanings, already accumulated knowledge. If you did not know anything about this 1944 BBC broadcast and only learned about it from this article, try going back and reread a few previous paragraphs of this article. Repeated reading will lead to a completely different meaning.

Unfortunately, "measuring of meaning" is impossible. Moreover, the point here is not so much in its subjectivity as in that any attempt to "measure" leads to a change in the initial meaning. The situation here unambiguously resembles the situation with the quantum measurements problem. It is well-known that a quantum system cannot be measured without changing its state. The same can be said when setting up an experiment to measure meaning. Such experience presupposes the "bombardment" of a person with questions, each of which requires comprehension and, as a result, the incorporation of the question into the integral meaning of the demiurge.

That is a typically quantum effect (described in detail in the works of Mensky, in particular, in the monograph [4]), which, however, manifests itself in the macrocosm. Unfortunately, no device would be such a delicate instrument that the subject would not undergo any changes during the experiment. Even bombardment with the most innocent questions – questions considered by the subject almost as nonsense – leads to an irreversible change in its integral cumulative meaning. In quantum measurements, it is even possible that the properties discovered during the measurement may not exist at all before the measurement, and the measurement of meaning can also lead to similar paradoxes. Measuring meaning can also lead to similar paradoxes.

Every opinion, every connotation has meaning only in relation to some point of reference. For example, before the outbreak of World War II, few people had a negative opinion of fascism. Indeed, at that time, fascism was a new, popular, and promising invention of the progressive socialists. The solutions fascists proposed to the problems faced by most countries took some people's breath away.

Some countries on both sides of the Atlantic began to implement reforms based on fascist ideas. However, this fascination with fascism turned out to be short-lived, and by the middle of the 40s, world public opinion of fascism radically changed from positive to negative. Humanity had reached a new level of understanding of this phenomenon relative to which fascism was far below.

For syntactic information, the binary concepts "true" and "false" are often introduced. Unfortunately, this is incorrect – information (in the technical sense), according to Shannon, does not contain any paradigms such as "true" or "false."

Only a subjective interpretation of the information received can lead to the emergence of the categories of "truth" and "false." Simultaneously, even "false information" can lead to some meaning different from the level of nonsense. For example, communist propaganda in the Soviet Union threw a tremendous amount of disinformation into society. Still, most of the population quickly learned to "read between the lines" – to fish out the grain of truth in a stream of lies – and thus developed a kind of immunity to propaganda. The same thing is happening now in other totalitarian countries.

We must put the familiar concepts of "truthful information," "false information," and "disinformation" in quotation marks because these terms, although widespread, are incorrect. The use of these terms simply means that the semantic interpretation of some information will be (subjectively) assessed either as having a non-trivial meaning or as nonsense.

At the same time, the category "truth," of course, exists. By "truth," we mean slowly changing (on a scale of several generations) geometric mean (for the ensemble of all humankind) meanings. For example, the fact that the Earth is flat was an indisputable truth several centuries ago. At present, humankind's geometric mean on this issue (i.e., generally accepted dogma) says that the Earth is a sphere. Nevertheless, this is precisely the average – there are still people who perceive this as nonsense, including the author of this article. For example, some people still believe that the Earth is flat, and the author of this article belongs to those who believe that the Earth is not a sphere but an ellipsoid.

Generally, this is the dilemma of all generally accepted dogmas. Of course, they can turn into something static, unchanging, and immune to reconsideration and dissent – in other words, sacrosanct. However, passing them on from generation to generation is necessary for at least three reasons. First, a certain "truth" can eventually turn into "absolute truth," that is, the truth, which is a logical consequence of certain postulates and axioms. Second, as follows from Gödel's theorem, any system of axioms is incomplete. Consequently, thirdly – if only because of the hope that the generally accepted dogma will be revised in the end.

Note here the difference between nonsense and absurdity. Absurdity is an expression that contradicts "absolute truth," i.e., opposes the system of axioms. For example, a "square circle" is absurd but is not nonsense since the meaning of a "square circle" is to serve as an illustration of the violation of axioms. To use the classical terminology of Kant, expressions like "square circle" are analytically absurd.

Another significant difference between information and meaning is manifested in their dynamics. For example, the amount of information in the text is a constant value that is not subject to the influence of time (it is assumed, of course, that the physical carrier of information is stable and not subject to erosion). Nevertheless, the meaning of what is written is continually changing – it changes rather slowly during one generation's lifetime, but after several generations, the meaning ascribed to any text changes.

The degree of such a change differs significantly for different texts. Still, the dynamics here are unambiguous – as a rule, over sufficiently long periods, the meaning of any text decreases, and in the limit reaches a unitary level, that is, the level of nonsense. A typical example is the Egyptian hieroglyphs, which were incomprehensible graphic symbols for millennia after the collapse of the ancient Egyptian civilization. It was clear that the hieroglyphs mean something, but what exactly was unknown. The meta-information that hieroglyphs exist carried a minimal, unitary, level of meaning until Napoleon's expeditionary force discovered the Rosetta Stone near Alexandria. As a result, Champollion deciphered the hieroglyphs, restoring the age-old traditions of relaying the transmission of meanings.

However, one paradox should be noted related to information and meaning. For example, if no one opens a book that has merely been on the bookshelf in the university library for centuries, then the amount of information in it does not change. However, that is not true for meaning.

*The meaning of what is written in any book changes, even if no one reads it.*

That, at first glance, the paradoxical conclusion is a direct consequence of the fact that words are only carriers of meaning but are not meaning as such. Indeed, the information is inside the book, and the meaning is outside it because the information is objective, and the meaning is subjective. As a part of human culture, language – as a method of encoding meanings into certain symbols – is continuously changing. The meaning changes over time because the language itself changes. The metamorphoses of language occupy an extensive range – what was considered nonsense yesterday may become something that is generally accepted tomorrow, and vice versa.

The peoples who created a particular language can leave the historical arena, and the language becomes extinct. Everything that was written in this language "will lose all meaning." Perhaps there will remain the experts of this dead language somewhere, but all other people will be able to get acquainted with the meaning of the written long-ago texts only with the help of translating ancient manuscripts into any modern language.

### 4. Thermodynamics of meaning

These simple considerations make it possible to postulate several laws related to meaning dynamics by directly borrowing from classical thermodynamics.

*The first law of thermodynamics of meaning: Perpetuum Intellectus is impossible.*

Intellectus in Latin means meaning, understanding, knowledge. The term Perpetuum Intellectus (a meaning that exists infinitely by itself, without any efforts of humanity) is introduced here by analogy with the term Perpetuum Mobile (a machine that works endlessly without an influx of external energy).

In other words, since the meaning of what is said and the meaning of what is written has a tendency to irreversible dissipation, then certain (sometimes even significant) efforts of society are required in order to transmit to the next generations not only information in the form of some long-lasting symbols but also the meaning that these symbols carry.

*The second law of thermodynamics of meaning: the entropy of meaning – as a measure of the chaos of meaning – increases in a closed system.*

An example of closed systems can be, for example, a closed book or any other storage medium that is not used for its intended purpose. In the extreme case of a dead language, the meaning is entirely suppressed by the chaos of incomprehensible symbols. Therefore, as a rule, human civilization's efforts are focused on methods of reducing the entropy of meaning, that is, establishing a specific (and accepted by the overwhelming majority) order in the methods of encoding meaning with graphics or other symbols.

It is interesting to note that even if the entropy of meaning in a closed system can change over time, the informational entropy – Shannon's entropy – remains constant.

Another widely used technical term, signal-to-noise ratio, can also be useful in describing meaning dynamics. An increase in entropy means that the useless noise begins to prevail over the useful signal, and the meaning "drowns" in the noise. The efforts of civilization to reduce the entropy of meaning mean that if noise is suppressed with a certain energy and financial costs, then the useful signal (meaning) is clearly detected.

One way to reduce the entropy of meaning is to translate from one language to another. In addition to the simple transfer of information from speakers of one language to speakers of another language, the translation process does not allow the irreversible dissipation process to destroy the meaning. With correct translation, the original meaning will pass without change to another language; only the graphic symbols for its display will change.

A good illustration of this is the well-known fact that the children of immigrants in America, in whose families they speak their native language (and English in school), as a rule, are ahead of their peers in development. That is why classical school education in Europe (many years ago) included an in-depth study of foreign languages, including one of the dead languages – Latin.

*The third law of thermodynamics of meaning: in the event of the death of human civilization (that is, in the extreme case, when the energy of society, aimed at maintaining the transfer of meaning from one generation to another, becomes zero), the entropy of meaning takes on a zero value, which ultimately leads to complete dissipation (destruction) of meaning.*

It should be noted that in this limiting case, the information, if it was recorded on reliable media, can be completely preserved, but the meaning completely disappears. That is precisely the meaning of information-semantic eschatology.

## 5. Relativity of meaning

In addition to directly borrowing from thermodynamics, the evolution of meaning can be described from the standpoint of another branch of physics – the theory of relativity.

The meaning of any text should be considered only from the point of view of the language (that is, the method of encoding thoughts) in which it was written. Thus, it makes sense to consider the meaning only within the framework associated with its original, own frame of reference. This frame of reference is a set of more or less strict rules for encoding meaning using symbols that existed at the time the text was written.

That is why the Catholic Church has always had a negative attitude towards church services not in Latin, and Muslims do not recognize the Koran, translated from Arabic into other languages. In both cases, it was merely a method by which attempts were made to transmit meaning unchanged from generation to generation. However, in both cases, this led to disappointing results – Latin became a dead language, and the Koran significantly slowed down the development of the Arabic language and, thus, Arabic culture.

Problems begin when humanity's frame of reference moves away from the frame of reference of the written text either at a sufficiently large distance (which is inevitable) or at a small distance, but very fast. At the same time, we postulate that the speed of dissemination of meanings in society is much lower than the speed of dissemination of information.

If society does not make any efforts to continually translate the text from the ancient language into the modern language, the meaning of what is written over time is guaranteed to fall to the level of nonsense, no matter how slowly the language itself changes. In this case, the meaning "will be left far behind."

For example, let us compare several versions of the New Testament. For example, the same phrase from *Hebrews 1:1* looks like this:

| **English Standard Version, 2001** | "Long ago, at many times and in many ways, God spoke to our fathers by the prophets" |
|---|---|
| **The New Jerusalem Bible, 1985** | "At many moments in the past and by many means, God spoke to our ancestors through the prophets" |
| **American Standard Version, 1901** | "God, having of old time spoken unto the fathers in the prophets by divers portions and in divers manners" |
| **King James Version, 1611** | "God, who at sundry times, and in diverse manners, spake in time past unto the fathers by the prophets" |

| **Early Wyclif Version, 1382** | "Manifold and many manners sometime God speaking to fathers in prophets" |
| **Latin Vulgate, around 405** | "Multifariam, et multis modis olim Deus loquens patribus in prophetis" |

Obviously, numerous attempts were made to convey the meaning of what was written throughout two millennia, using a language understandable to contemporaries. That is why there is not one but many versions of the Septuagint, the Greek translation of the Old Testament. The Shannon entropy changed from version to version, the amount of information changed, but the meaning of what was said did not change. Of course, the loss of some nuances during translation happens – not without it – but the primary meaning is never lost at long enough time intervals. Later, we will consider an example of how an unscrupulous translation can lead to the loss of some nuances and grave consequences.

The frame of reference in which the meaning was encoded into graphic symbols always remains in the past, and each new generation is simply obliged to adapt ancient texts to modern language. Otherwise, the meaning will slide down to the level of nonsense.

Sometimes this adaptation takes the form of harsh political confrontation.

For example, in America, there are two leading schools of constitutional jurisprudence. One school argues that the U.S. Constitution should be viewed in the context of the language and culture that existed at the time of its writing at the end of the 18$^{th}$ century. Supporters of this school argue that adaptation to a modern language is necessary, but only without losing the original meaning. As supporters of this school prefer to say, it is necessary to interpret the Constitution from the standpoint of "as it was written." This position is usually called originalism, or the doctrine of the original meaning.

The alternative school argues that the U.S. Constitution is a "living document" and suggests "reverse adaptation." They propose to interpret the Constitution not by building a semantic "bridge" between the old and modern languages but by adapting the language of the Constitution to the current political realities by replacing "old" terms with "new" ones.

In other words, instead of assigning old meaning to new, modern words, the proponents of this school insist on assigning new meaning to old words.

Of course, this approach leads to the Constitution's actual change and gives it a new meaning, which may coincide or may contradict the opinion of the Founding Fathers of the United States. Actually, these two schools' battle boils down to the fact that one school proposes to reread the U.S. Constitution continually, and the other – to continually rewrite it.

Extrapolating this dichotomy to any other source of information, one can say that only rethinking (rereading) can act as a mechanism for transferring knowledge from one generation to another. An alternative to this is an adaptation (rewriting) – a process that, as a rule, leads to an increase in the entropy of meaning.

Of course, the phenomena described here are not new. Confucius, back in the 4[th] century B.C., suggested "name correction" to "bring words in line with reality." According to him, "If names are not correct, language is not in accordance with the truth of things. If language is not in accordance with the truth of things, affairs cannot be carried on to success. "[8]

An excellent example is the use of the terms "conservatism" and "liberalism." At present, the meaning attributed to them, in most cases, has nothing to do with the original meanings of these words.

Initially, the word "conservatism" meant exclusively British conservatism, in contrast to which British liberalism was created (mainly by the British and French philosophers) in the 18[th] century. The main idea of liberalism is that, within the man-state paradigm, supremacy should belong to the man. The Founding Fathers of America were excellent students and put into practice the theoretical ideas of liberalism on American soil.

For the first half of its existence, the United States called them so – liberals, but later this term was forcibly redefined, and completely different people, ideologically very far from liberalism, appropriated this term for themselves. Those who remained faithful to freedom's achievements were called "conservatives," and those who usurped the term liberalism were called "liberals."

As a result, people come into existence in America who, to the horror of Europeans, are both conservatives and liberals at the same time. How this became possible in America is still a headache on both sides of the pond. Many refuse to understand that the American conservatives are the actual, original liberals, and the current liberals have nothing to do with liberalism. Even 26 centuries ago, Confucius warned humanity about the danger of words that are not brought in line with reality, but people still prefer to step on the same rake.

At present, both American conservatism (in Roger Scruton's terminology, "empirical conservatism") and British conservatism ("metaphysical conservatism") appear under the same name – conservatism. Nevertheless, if British conservatism is to be understood in a static sense (the inviolability of the status quo), American conservatism should be understood in a dynamic sense (inviolability of constitutional freedoms).

Unfortunately, bringing words into line with reality – rethinking (rereading) – never leads to discovering something new. For humanity's progress, old meanings are necessary and their alteration up to the abolition of dogmas, generally accepted meanings, and "written in stone" traditions – only in this way can new meanings arise.

*Thus, human civilization's overall progress is based on the dual process of information rereading-rewriting.*

This dualism is manifested in the fact that humanity is simultaneously working both to reduce the entropy of old meanings (making the already known knowledge crystal clear and consistent) and to increase the entropy for new meanings (only heuristics – the introduction of certain doubts, uncertainties, and refusal from old dogmas can lead to the discovery of new knowledge).

The ideal would be a situation of equilibrium – when the decrease in entropy during rereading would be fully compensated by the increase in entropy for rewriting (for scientific research).

However, as we know, humanity is very far from such an equilibrium state. At the same time, this dual process also manifests itself at the individual's level – rereading prevails during sleep, and rewriting prevails during wakefulness. From this point of view, sleep is a necessary condition for the existence of consciousness.

As a rule, changes in both the language and the texts' meanings are slow; more precisely, they occur equally slowly. Nevertheless, everyone knows situations when a language change did not lead to a corresponding change in meaning and vice versa. This effect occurs when the time frame of the process is small (one generation, for example), but the rate of changes either in the language itself or the meaning it carries is significantly higher than usual.

For example, during the French Revolution at the end of the $18^{th}$ century, the language was changed. Changed forcibly – the revolutionaries changed the months' names, canceled the weeks and renamed the days, changed the hours (the day became 10 hours for 100 minutes, etc.) Has the meaning changed? No, the people continued to use the "old calendar." As a result, the new system was canceled as quickly (in historical terms) as it was introduced. The violent change of language was ignored.

Another example is associated with the communist coup in Russia at the beginning of the $20^{th}$ century. The communists seized power and almost immediately began to modify the semantic load of the Russian language. In their opinion, the new Russian language had to reflect the new communist ideals, and as quickly as possible. Unlike the French leftists, the Russian leftists began not with language but with an artificial redefinition of concepts, and only then brought the Russian language into line with the new realities. The people were forced first to use new terms, then the old ones, which, however, were forcibly assigned an alternative meaning, and then they brought down new rules of grammar on the citizens.

Of course, such a dictatorial attitude towards one's language should have led to an increase in the entropy of meaning, and it happened. The regular transmission of the relay of information and the relay of meaning from generation to generation was interrupted, but the communists never viewed this process from a negative point of view. On the contrary, the destruction of everything "old" was welcomed, and within one or two decades, another Russian culture and another Russian language, the forerunner of Newspeak, was created.

In contrast to the French left's dictatorship, the consequences of the communist dictatorship are still present in modern Russian. However, the most alarming is the fact that an example of a "successful" breaking of the adaptation mechanism (technically, this was done with the help of physical destruction of the carriers of the "old" culture, total censorship, and total terror) still haunts those who would like to repeat it.

For example, in the United States, an artificial attempt to substantially redefine the meaning while leaving the language structure practically unchanged, creating an alternative culture, took the form of political correctness. Terms have appeared that carries an entirely different meaning, although they are expressed in familiar words – for example, "climate change." Nevertheless, the artificial endowment of the term "climate change" with anthropogenic content will be fixed in the language only in the unlikely event that a transition to some form of dictatorship is made in America. This applies to all attempts to manipulate people's thoughts and views through deliberate language changes.

## 6. Evolution of information and evolution of meaning

The immutable fact that the viability of meaning depends entirely on its carrier of information should be noted. Outside the carrier, meaning does not spread beyond one person; therefore, the dynamics of meaning depend not only on the subjective extraction of meaning by people but also on the objective, physical change of the information carrier.

If the information carrier is long-lived, then the dynamics of adaptation of language and meanings slows down because humanity has no incentive to change anything. Vice versa, if the information carrier is short-lived, then humanity is forced to make every effort to ensure the timely synchronization of language and meaning.

For example, rock paintings and inscriptions carved in stone have been preserved for thousands of years, and the evolution of human culture in general (and language in particular) proceeded slowly. The process drastically accelerated when papyrus, and then parchment became the carrier. With the invention and mass adoption of paper – which is undoubtedly a much shorter-lived medium than stone and leather – the dynamics of culture and language have accelerated significantly.

The digital revolution has led to an even more revolutionary era (there is simply no other suitable word other than revolutionary), in which the duration of information storage ceased to depend on the carrier – it began to depend only on the person. At the person's request, the storage duration can be zero (if the person decided to delete the newly created file) or extend for many billions of years (if a nickel disk acts as a carrier).

In the digital world, there is a remarkable and well-known phenomenon. Due to the excess of information, people often simply cease to extract meaning from it. At the beginning of the $20^{th}$ century, families listened to every announcer's word on the radio to not miss something important. At the beginning of the $21^{st}$ century, the overwhelming majority of information is "not received." It is filtered out. Not only is it not remembered, but even no attempt is made to derive any meaning from the flow of information. In other words, in the digital era, the "meaningfulness" (*intellecaptus*) of information (the amount of subjective meaning that a unit of information carries) falls significantly.

As a result of a selective approach to extracting meaning from the information flow, information sources in modern society are rapidly segregated. It is reflected in the well-known paradigm, expressed by Paul Simon as "A man hears what he wants to hear, and disregards the rest." That has led to an intense polarization of information sources and the meanings people extract from them.

As a result, people from one cohort "do not hear" people from another cohort in no small measure because they "disconnect" themselves from those sources of information they do not trust (i.e., news nihilism). The point is not that one cohort "does not understand" or "understands, but refuses to accept" other points of view, but that a group of people physically disconnects themselves from information flows, which they consider unacceptable. At the same time, as it is known, the best thinkers know and understand the points of view that are opposite to their convictions, although they do not share them.

In practice, the described effects appear in the form of the so-called "news cycle," when the audience is receptive to news only for a short period of time, and then "getting used" to the news, and begins to pay attention to them only after the next "sensations" (real or fictional – does not matter).

Based on this, several years ago, the author proposed the concept of "news half-life." The half-life of news is approximately 24 hours; that is, the next day, about half of the listeners (or viewers) forget what was in yesterday's news. Two days later, half of the remaining half of the audience also forgets about the sensation. So, after two days, only a quarter of those who remember something remains, and so on.

The process of transferring old knowledge (old meanings) to new generations often resembles a professor's work with those students who did not understand the new material right away. In this case, the professor is forced to explain complex material to such students over and over again, each time using some new techniques, until, finally, his arguments (his carriers of information) lead to the fact that the student himself will build a "bridge" between new knowledge (new meaning) and old knowledge (old meanings).

### 7. Distortions of meaning

All the above referred to the cases of the "ideal" process of extraction and adaptation of meanings. At the same time, humanity knows numerous examples of how the adaptation took place in bad faith. As a result, the meaning of ancient manuscripts was transmitted with significant distortions.

An example of an unscrupulous approach is the famous passage from the Gospel of *Matthew 19:24*, which says that "it is easier for a camel to go through the eye of a needle than for a rich person to enter the Kingdom of God." The most likely explanation seems to be that the translator from the Aramaic, apparently, not only did not live in Jerusalem but was not familiar with Jerusalem, which had a (very narrow) gate under the unofficial name "Eye of a needle." As a result, this incorrectly translated phrase led to numerous, two thousand years of discussions and hundreds of volumes of unnecessary theological research. This ridiculous phrase is often attributed to some hidden, deep, and mystical meaning, which was not and could not have been in the original (oral) phrase.

In addition to the "external" causes of distortions of meaning, there can also be "internal" ones. It happens due to a malfunction in the brain (for example, due to ordinary forgetting). As a result, the already accumulated meaning could be distorted.

If the given examples can be attributed to the category of "innocent" mistakes in the adaptation of meanings, then the following examples are much more severe. One of them led to anti-Semitism and the Holocaust.

The first episode of deliberate distortion of the Biblical meaning is relatively recent and is associated with the Chinese Communists. In their version of the New Testament, the episode with the harlot (everyone knows the episode in which Jesus says: "He that is without sin among you, let him be the first to throw a stone at her") ends with Jesus himself throwing stones and killing the harlot. The political underpinnings of such falsifications need no comment.

Before citing another example, we note that all the canonical (but not apocryphal) books of the New Testament have come down to us in Greek, not in Aramaic (Aramaic was a spoken language). The translators were native speakers of the language – the Greeks.

After Greece's capture by Rome in 146 BC, the Greeks in the Roman Empire occupied a privileged place – the intelligentsia place. The profession of pundits that they occupied required a wise and dodgy policy towards the Empire. Like Judea, Greece was occupied by Rome, and the conquered Greeks fought in every way for their privileges – usually at the expense of other conquered peoples.

One well–known example of the privileged position of the Greeks was the incorporation of the Greek gods of Olympus into the Roman Pantheon. All history textbooks claim that "the Romans took over their gods from the Greeks," but this is not entirely correct. The Romans had merit in this, but it was minimal. The fact is that it was the Greeks, not the Romans, who incorporated the Greek gods into the Roman Pantheon. From the point of view of the theory considered here, this incorporation was made almost perfectly – when translated into Latin, the Olympian gods received new names (Zeus became Jupiter, for example), but the meaning remained almost the same, close to Greek mythology as much as the Romans would tolerate.

The Greeks' strategy concerning Rome consisted mainly of praising Rome, Caesar, Roman culture – and belittling all occupied peoples (except Greek culture, of course). The Greeks' strategy was successful – they managed to save their culture, their language, and then, after the Roman Empire's division, even restore their statehood in Byzantium.

The canonical versions of the New Testament in Greek bear all the signs of a Greek survival strategy – any actions of the Romans are exalted, and all the negativity falls on the Jews. It turns out that it was the Jewish Sanhedrin who sentenced Jesus to death. It turns out that Pontius Pilate (one of the cruelest procurators in Rome's entire history) made every effort to ensure that Jesus escaped death. It turns out that the Jewish high priests and elders went against the noble Pilate and insisted on the approval of the death sentence. The reader was unambiguously hinted that it was the "Jews who crucified Christ."

However, the realities of Rome two thousand years ago were quite different. There was simply no need for Pilate to stage the show described in the Greek version of events. No one can say with certainty which events in the canonical Gospels are historical truth. Nevertheless, by now, it has become clear that the nuances introduced into the translation by the Greeks completely rearranged the accents and significantly distorted the real history of Jesus Christ. The result of this dishonesty is well known – mass anti-Semitism and the Holocaust.

However, the story does not end there. Six centuries after the "noble Pontius Pilate" and the "evil Jews," elements of the Old and New Testaments were incorporated into Islam. Muhammad's interpretation not only upheld the anti-Jewish emphases but also reinforced them. If the anti-Semitism of the canonical Gospels was "hidden," then Muhammad legalized it.

In 1965, the Second Vatican Council adopted a document that, generally speaking, should have been adopted two millennia ago – Nostra Aetate [9]. It finally removes the collective guilt for the death of Jesus Christ from the Jews and condemns anti-Semitism. Israel's recent successes in the international arena are a direct result of this bold decision by the Vatican.

## 8. Conclusion

This paper omits such essential issues as the mechanism of transformation of information into meaning (hermeneutics), as well as the meaning of meaning – its ontology. Nevertheless, even leaving aside these issues (at the micro-level), consideration can lead to significant macro-level observations. In essence, the analogy with classical thermodynamics is unambiguous here. It is based on the fact that a large ensemble – either classical particles or meanings carried out by neurons – can under certain conditions be described without going into detail at the micro-level.

The statistical evolution of meanings and the intellectual level obtained at the macro level are in good agreement with phenomenological observations – at least qualitatively. Such a "physical" approach should turn out to be very fruitful because physicists, as a rule, (and unlike philosophers) never deal with solving all world problems at the same time but build a particular model of a specific process under study.

Physicists are well aware that they operate with models and only models. Nevertheless, this does not prevent physicists from applying mathematical statistics & methods to many objects' ensembles, even if the objects themselves' have an inner nature that is still mostly unclear.

The theory of meaning proposed here is "physical" in the sense that it is only a limited (by definition) model of the dynamics of meaning as a statistical ensemble of those phenomena that occur in the human mind.

The development of all aspects of this model is a matter of the future. Here are just a few problems. For example, building a model for meanings obtained without outside influence – intuition (based on already obtained meanings – this article can serve as an example), and considering the correlation effects between meanings conveyed in this way and meanings obtained by extracting meanings from external information.

A question also arises about the external parametric (with a period of 24 hours) excitation (modulation) in extracting meanings. In this case, about 2/3 of the time, the consciousness is mainly rewriting, and about 1/3 of the time – mainly rereading.

Finally, let us note the essential consequences of the proposed approach for the problem of artificial intelligence. Any artificial intelligence apparatus should have the same qualitative characteristics as human intelligence. That is, multiplicativity, non-commutativity, and irreversibility (that is, the impossibility of "taking away" the already obtained knowledge) should be the basis of any device that claims to be called "artificial intelligence."

**Appendix I.**

Let's introduce the cumulative meaning $M$, which at the step $N$ has the form

$$M_N = \prod_{i=1}^{N} m_i , \quad (1.1)$$

where $m_i \geq 1$ is the meaning obtained in step $i$, and $m_1 = 1$. For the cumulative meaning (knowledge) $M$, we also introduce the level of intelligence as a geometric mean:

$$I_N = M_N^{1/N} \quad (1.2)$$

For $N \gg 1$ the level of intelligence takes the form

$$I_{N+1} \approx I_N \left( \frac{m_{N+1}}{I_N} \right)^{1/N} \quad (1.3)$$

Or, expanding (1.3) in powers of a small parameter $(1/N) \ll 1$,

$$I_{N+1} \approx I_N \left( 1 + \frac{1}{N} \ln \frac{m_{N+1}}{I_N} \right) \quad (1.4)$$

The meaning of the expression (1.4) boils down to the fact that the level of intelligence at each step of comprehending the information received is directly proportional to the level of intelligence already achieved. The increase or decrease in intelligence level depends on the ratio of newly obtained meaning $m_{N+1}$ to the average intelligence level $I_N$.

If the knowledge received exceeds the current level of intelligence, then the logarithm in expression (1.4) will be greater than 0; that is, the intelligence level will increase. In the opposite case, if the newly obtained knowledge is less than the level of intelligence, then the logarithm in (1.4) will become negative; that is, a step will be taken towards degradation.

Consequently, an increase in the level of a person's intelligence occurs only when the new knowledge exceeds some already achieved level of intelligence for this person. If the new knowledge is small in comparison with the achieved level of intelligence, then intellectual degradation occurs.

Formula (1.4) shows that cognition is a flywheel, spinning up of which must be continuously increased to avoid intellectual degradation.

*Thus, from the point of view of the intellect, life is a constant race with itself.*

In the extreme case, when new knowledge, a new meaning has a nonsense level, $m_i = 1$, then the cumulative meaning $M$ does not change, but the level of intelligence falls.

If the meaning is considered nonsense for many steps (in other words, nihilism), a rapid intellectual degradation occurs. Sometimes, such degradation occurs with people alone, imprisoned, or people who, due to censorship, are deprived of access to new information. Moreover, degradation does not depend on how the censorship is carried out – from the outside (state censorship), or internally, when a person himself deprives himself of access to information (for example, for some ideological reasons refuses to pay attention to alternative points of view).

One possible exception to the degradation rule comes from the recent paper [10], which demonstrates that isolation can also stimulate areas of the brain that control creativity and imagination – to a certain degree, of course.

Another interesting conclusion from formula (1.4) is that the contribution to intelligence level decreases when $N$ increases. That is why older people find it more difficult to comprehend new information than young people.

**Appendix II.**

Formula (1.1) can be represented in the recursive form

$$M_{N+1} = M_N m_{N+1},$$

or

$$\frac{M_{N+1} - M_N}{M_N} = m_{N+1} - 1.$$

If the step is set in accordance with the time $T/\tau = N \gg 1$, then one can go to continuous variables,

$$\frac{1}{M}\frac{dM}{dt} = \frac{m(t)-1}{\tau}, \qquad (1.5)$$

where $\tau$ is a certain characteristic time of the comprehension process.

The solution to equation (1.5) with the initial condition $M(0) = 1$ (i.e., *tabula rasa* condition) is

$$M(T) = \exp\left\{\frac{1}{\tau}\int_0^T [m(t)-1]dt\right\} \qquad (1.6)$$

Integral representation (1.6) once again shows that the accumulation of knowledge, that is, a non-trivial contribution to the integral (1.6), is possible only if the incoming information is not assigned a level of nonsense.

In reality, the obtained meanings $m(t)$ are a random process, not a deterministic function. Let us assume that there is a certain average value $\langle m \rangle$ that does not depend on time (brackets $\langle . \rangle$ will denote averaging over the ensemble).

Averaging (1.6) over the ensemble of all meanings, we obtain a simple expression for the mean logarithm of the cumulative meaning:

$$\langle \ln M(T) \rangle = \frac{T}{\tau}(\langle m \rangle - 1) \qquad (1.7)$$

Expression (1.7) confirms the intuitive result that the accumulated meaning is proportional to both the time of accumulation and the average level of an excess of meaning over the level of nonsense. In addition, from expression (1.7) follows the well-known fact that "slow-witted" people with a considerable value of information processing characteristic time $\tau$ accumulate less knowledge during time $T$.

Expression (1.7) can be represented as

$$\langle m \rangle - 1 = \frac{\tau}{T} \langle \ln M(T) \rangle \qquad (1.8)$$

If we compare expression (1.8) with the definition of the Lyapunov characteristic exponent,

$$\lambda = \lim_{T \to \infty} \frac{\tau}{T} \ln M(T), \qquad (1.9)$$

then it can be argued that for the process of accumulation of meanings

$$\lambda = \langle m \rangle - 1 \qquad (1.10)$$

In the case that for a particular cohort of people, the average meaning exceeds the level of nonsense, i.e., $\langle m \rangle > 1$, then the Lyapunov exponent $\lambda > 0$ and specific realizations of the "trajectory of cognition" for different people will exponentially "diverge" even though the initial "tabula rasa" condition is the same for all people. For people from this cohort, there will be significant intellectual inequality.

The divergence of trajectories is an integral characteristic of unstable irreversible non-linear dynamic systems. If we add to this the irreversible dissipation of meanings (forgetting), there are all the known conditions for the stochastic self-organization of meanings.

For a cohort of people with $\langle m \rangle < 1$, specific realizations of the "trajectories of cognition" will "converge" (in the statistical sense, of course). It is often said about this cohort of people (people with low intellectual level) that "they all look the same," although there is an inevitable intellectual inequality among them too.

**Appendix III.**

Let us denote $m(t) = \langle m \rangle + \zeta(t)$, where $\zeta(t)$ is a delta-correlated random Gaussian process with a zero mean value, dispersion $\sigma^2$, and correlation function

$$\langle \zeta(t)\zeta(t') \rangle = 2\sigma^2 \tau \delta(t - t').$$

We rewrite solution (1.6) in the form

$$M(T) = \exp\left\{ \lambda \frac{T}{\tau} + \frac{1}{\tau} \int_0^T \zeta(t) dt \right\} \qquad (1.11)$$

The corresponding Fokker-Planck equation for the probability density $P(M,t)$ of the cumulative meaning $M$ at the moment of time $T$ looks like this [11]:

$$\left( \frac{\partial}{\partial t} + \lambda \frac{\partial}{\partial M} M \right) P(M,t) = \frac{\partial}{\partial M} M \frac{\partial}{\partial M} M P(M,t)$$

(1.12)

where $t = T/\tau$ is the dimensionless time.

The solution to equation (1.12) is a log-normal process with the probability density

$$P(M,t) = \frac{1}{2M(\pi \sigma^2 t)^{1/2}} \exp\left\{ -\frac{\ln^2[M \exp(-\lambda t)]}{4\sigma^2 t} \right\}$$

(1.13)

The process of accumulating knowledge with the probability density (1.13) is a Markovian non-stationary process. The distribution function (1.13) has a much longer "tail" than the Gaussian random process. That suggests that outliers of a random process $\zeta(t)$ play a disproportionately large role in forming cumulative meaning.

Let us note here another paradox. The described Markov process of accumulation of meanings is a process "without memory." However, knowledge, meanings are stored precisely in the memory of a person.

*Thus, from a mathematical perspective, forming a person's memory itself is a process without memory.*

In other words, in the described process of knowledge accumulation, "moments of enlightenment" play an essential role when understanding something complex suddenly develops in a person's mind. It is these "illumination of truth" moments that create the long "tail" in the distribution (1.13). In physics, such phenomena are called the intermittency effect, characterized by rare but significant in magnitude bursts.

The ensemble mean value of the cumulative meaning

$$\langle M(t) \rangle = \exp\left[\left(\sigma^2 + \langle m \rangle - 1\right)t\right] \quad (1.14)$$

grows exponentially over time. Note that this growth also occurs when the average value of the received meanings is nonsense, that is $\langle m \rangle = 1$. That happens because of the "spikes" of a random process of accumulation of meanings $\zeta(t)$, that is, from a small number, but significant in magnitude, "eureka moments."

The variance of the cumulative meaning also grows exponentially,

$$\langle M^2(t) \rangle = \exp\left[2\left(2\sigma^2 + \langle m \rangle - 1\right)t\right] \quad (1.15)$$

Moreover, in the case of $\langle m \rangle = 1$ the variance still grows. In other words, even among nihilists, intellectual inequality is guaranteed to be observed for a long enough time.

In the unique case when the condition

$$\sigma^2 + \langle m \rangle - 1 = 0 \quad (1.16)$$

is met, the average value $\langle M(t) \rangle = 1$ does not depend on time, but the variance still grows according to the law $\exp(2\sigma^2 t)$. Thus, even if $\langle m \rangle - 1$ is small, this is compensated by the variance of the received meanings.

In other words, even among non-clever people, or people with some mental disorders, geniuses can appear due to the term $\exp(2\sigma^2 t)$. It is easy to identify such people – they belong to the cohort of people whose average accumulated knowledge is simply nonsense.

Expression (1.15) once again proves that the "mentally retarded" people under no circumstances shall be written off. Anyone in doubt is encouraged to watch the *Rain Man* movie.

In continuous time, the level of intelligence (1.2) can be represented as

$$I = M^{\tau/T} = \exp\left[\lambda + \frac{1}{T}\int_0^T \zeta(t)dt\right] \quad (1.17)$$

Its average value

$$\langle I \rangle = \exp\left[\langle m \rangle - 1 + \frac{\sigma^2}{t}\right] \quad (1.18)$$

The exponent in expression (1.18) contains two terms, the first of which does not depend on time. This part of the intellectual level is due to the average level of the received meanings in relation to the level of nonsense $\langle m \rangle - 1$. The second term is due to fluctuations in the received meanings, but this term decreases rapidly with time.

So, the expression

$$I_{\min} = \exp(\langle m \rangle - 1) \qquad (1.19)$$

represents the lower bound of the average level of intelligence when $t \to \infty$.

Finally, a direct generalization of the results obtained is possible for a model with "forgetfulness." To take into account the fact that it is common for people to forget, in all the formulas of Appendix II and Appendix III, we make a simple substitution $\langle m \rangle \to \langle m \rangle - \beta$, where $1/\beta$ is the characteristic dimensionless "half-life of meanings," or "forgetting time."

**Acknowledgments**

The author is grateful to Prof. Oleg Butkovsky and Peter Miller for fruitful discussions; Dr. Igor Mandel and Dr. Stan Lipovetsky for a thorough review. Their constructive remarks made it possible to improve the structure of the article significantly.

**References**

[1] I.A. Richards, and C.K. Ogden, *The Meaning of Meaning*. Harvest/HBJ 1989

[2] Descartes, Rene: *Traite de l'Homme* (1664). Translated by T.S. Hall, *Treatise of Man,* Prometheus, 2003

[3] Penrose, Roger, *Shadows of the Mind: A Search for the Missing Science of Consciousness*. Oxford University Press, 1994

[4] M.B. Mensky, *Consciousness and Quantum Mechanics*, World Scientific Publishing Company, 2010

[5] A. M. Zheltikov, *The critique of quantum mind: measurement, consciousness, delayed choice, and lost coherence*, Phys. Usp., 61:10 (2018), 1016–1025

[6] *The Routledge Handbook of Philosophy of Information*, Edited by Luciano Floridi, Routledge, 2016

[7] Pribram, Karl. *Brain and perception: holonomy and structure in figural processing*. Lawrence Erlbaum Associates, 1991

[8] Legge, James, *Confucian analects: The great learning, and the doctrine of the mean,* Dover Publications, 1971

[9] Nostra Aetate, https://en.wikipdia.org/wi-ki/Nostra_aetate

[10] Spreng, R.N., Dimas, E., Mwilambwe-Tshilobo, L. et al. *The default network of the human brain is associated with perceived social isolation.* Nature Communications 11, 6393 (2020)

[11] V.I. Klyatskin, *Dynamics of Stochastic Systems*, Elsevier, 2005